\documentclass[twocolumn]{aastex63}
\usepackage{graphicx}
\usepackage{color}
\usepackage{amsmath}
\usepackage{natbib}
\usepackage{float}
\usepackage[caption = false]{subfig}
\usepackage{hyperref}
\usepackage{fancyhdr}

\newcommand{\be}[1]{\begin{equation} \label{eq:#1}}
\newcommand{\ee}{\end{equation}}
\newcommand{\ba}[1]{\begin{eqnarray} \label{eq:#1}}
\newcommand{\ea}{\end{eqnarray}}

\newcommand{\solrad}{\ifmmode{R}_{\rm S}\else${R}_{\rm S}$\fi}
\newcommand{\solmas}{\ifmmode{M}_{\rm S}\else${M}_{\rm S}$\fi}

\newcommand{\tintu}{\ifmmode{\rm erg~cm^{-2}~s^{-1}sr^{-1}}\else 
  erg~cm$^{-2}$~s$^{-1}$~sr$^{-1}$\fi}
\newcommand{\fluxu}{\ifmmode{\rm erg~cm^{-2}~s^{-1}}\else 
  erg~cm$^{-2}$~s$^{-1}$\fi}

\newcommand{\wave}{\ifmmode{\lambda} \else$\lambda$\fi}

\newcommand\lta { \mathrel {\hbox to 0pt {\lower 3.7pt \hbox{$\sim$}
      \hss} \raise 1.7pt \hbox{$<$}}}
\newcommand\gta { \mathrel {\hbox to 0pt {\lower 3.7pt \hbox{$\sim$}
      \hss} \raise 1.7pt \hbox{$>$}}}

\begin{document}

%
%

\title{GOES class estimation for behind-the-limb solar flares using MESSENGER SAX}

\submitjournal{The Astrophysical Journal}

A modified version of this paper was accepted for publication in The Astrophysical Journal.


\correspondingauthor{Erica Lastufka}
 \email{erica.lastufka@fhnw.ch}
 
\author[0000-0003-1894-2074]{Erica Lastufka}
\affiliation{Fachhochschule Nordwestschweiz,\\
       Bahnhofstrasse 6, 5210 Windisch, Switzerland}
\affiliation{ETH-Z\"urich,\\
       R\"amistrasse 101, 8092 Z\"urich, Switzerland}

\author[0000-0002-2002-9180]{S\"am Krucker}
\affiliation{Fachhochschule Nordwestschweiz,\\
      Bahnhofstrasse 6, 5210 Windisch, Switzerland}
\affiliation{Space Science Laboratory, UC Berkeley}



%
%

\begin{abstract}
Mercury mission MESSENGER's 
Solar Assembly for X-rays (SAX) observed almost 700 solar flares between May 28, 2007 and August 19, 2013, as cataloged by \citet{dennisSOLARFLAREELEMENT2015}. The SAX instrument, part of the X-ray Spectrometer (XRS), operated at 1 -- 10 keV, partially overlapping the energy range of the GOES X-ray spectrometers. SAX provides viewing angles different from the Earth-Sun line and can therefore be used as a GOES proxy for partially or fully occulted flares as seen from Earth. 
For flares with GOES classes above C2 seen on-disk for both instruments, we found an empirical relationship between the soft X-ray (SXR) fluxes measured by both SAX and GOES. 
Due to the different energy response of the two SXR instruments, individual events can deviate on average by about a factor of two from the empirical relationship, implying that predictions of the GOES class of occulted flares from SAX data are therefore accurate to within the same factor.


The distinctive GOES energy response in combination with the multithermal nature of flares makes it difficult for any instrument, even other soft X-ray spectrometers, to provide a GOES proxy more accurate than a factor of two.

\end{abstract}

\keywords{Flares, Soft X-rays, Extreme ultraviolet, GOES, MESSENGER, STEREO}

\section{Introduction}

Since 1969, solar flare classifications have been made using the Geostationary Operational Environmental Satellite (GOES) series of satellites operated by the National Oceanographic and Atmospheric Administration (NOAA) \citep{bakerFlareClassificationBased1970}. 
Soft X-ray-sensitive detectors of the X-Ray Sensor (XRS) measure the total flux in two wavelength bands -- long wavelength (1--8 \AA{}) and short wavelength (0.5--4 \AA{}). 
The GOES class of a flare, ranging on a logarithmic scale from A (10$^{-8}$ $W/m^2$ ) to X ($\geq$ 10$^{-4}$ $W/m^2$), is given by the value of the peak flux in the GOES long channel during a flare event. 

This method of solar flare classification was adopted for its advantages over the optical classification in determining a flare's geophysical importance. The choice of the 1-8 \AA{} band in particular was driven by the fact that radiation at those wavelengths was thought to drive ionospheric disturbances \citep{donnellySolarActivityReports1979}. Although its origins lie in spaceweather forecasting, the GOES classification system has been adopted by the larger solar community as an indicator of a flare's magnitude. 

However, it is worth asking ourselves what the peak value of disk-integrated soft X-ray flux physically indicates. First of all, during most large flare events, the GOES SXR signal from both solar and non-solar background sources will be in the percent range compared to the signal from the flare. For the identification of smaller flares, especially during solar maximum, or for situations where the peak of one flare overlaps with the decay of another, background subtraction becomes non-negligible \citep{ryanThermalPropertiesSolar2012}. Therefore it is possible that the regions outside the flare of interest contribute to the peak 1--8 \AA{} flux used to determine event GOES class.

Secondly, we must consider the geometry of the X-ray emitting region. A single value of area-integrated flux can just as easily represent a compact source as an extended one. 

Consider two isothermal flare plasmas of equivalent temperature and thermal energy, but different volumes. Equal energies means an equal number of particles $N$ ($E_{th} = 3NkT$, where $T$ is the temperature and $k$ the Boltzmann constant, \citep[e.g.][]{hannahMicroflaresStatisticsXray2011}); therefore, one plasma must have a higher density than the other. Emission measure is dependent on both density and volume ($EM = n^{2}_{e}V$ = $NV^{-1}$), meaning that ultimately it is inversely proportional to the volume $V$. If one plasma has a spherical volume with radius 5" and the other has a radius of 50", this results in the compact plasma having an emission measure one thousand times higher than the extended source, and thus having a GOES classification three orders of magnitude higher, despite the two plasmas having the same thermal energy content!

Fortunately, soft X-ray-emitting flare plasmas are just as spherical as chickens; flare loops are the main source of SXR emission. In any individual loop, the volume will then be given by a long loop length and a relatively much smaller cross section. This makes  the volume dependence essentially a length dependence, and a much more reasonable factor of 10 versus factor of 1000 is recovered for the aforementioned plasmas. Still, it is easy to see how GOES class can seem biased towards more compact sources. It is important to keep in mind that GOES class is not an accurate indicator of the total energy or energy release potential of a flare.

Nevertheless, the GOES classification scheme is universally accepted and commonly used as a shorthand for flare size as well as a selection criterion for statistical studies. 
Many properties of flares, most importantly temperature, emission measure, non-thermal energy, and total radiative energy losses have been shown to scale with GOES class, especially for flares classified at GOES M or above \citep[e.g.][]{warmuthConstraintsEnergyRelease2016}.

GOES satellites are in a Sun-synchronous orbit and as such cannot be used to classify events that happen behind the solar disk as seen from Earth. Other Sun-observing instruments can provide this perspective; the most well-known are the twin STEREO satellites \citep{howardSunEarthConnection2008}, whose Sun-Earth Connection Coronal and Heliospheric Investigation (SECCHI) suite provides imaging in multiple extreme ultraviolet (EUV) channels. STEREO provided complete coverage of the Sun from 2007 until the Behind satellite stopped operating in 2014; however, even a partial view of the Sun's far side provides valuable information about flares and coronal mass ejections (CMEs) that originate from active regions hidden to the Earth-orbiting view. 

The wealth of STEREO data made it especially advantageous to find a relation between EUV observations and soft X-ray flux, in order to provide a GOES class estimate for behind-the-limb flares.
\cite{nittaSoftXrayFluxes2013} derived an empirical relationship between EUV fluxes and the GOES class. 
\cite{chertokSimpleWayEstimate2015} took a different approach, using the length of the horizontal saturation streaks that appear in STEREO 195 \AA images to derive a GOES class. STEREO data is inherently limited by its five-minute cadence, making timing critical, and the Chertok method is additionally confined to events that are EUV-bright enough to produce a saturation streak in the first place. Both sets of authors reported better fits to the data for flares with higher GOES class, with Nitta quoting an uncertainty of under a factor of 2 for large flares. 
However this is not unexpected, as the double-peaked response function of the 195 \AA{} channel clearly differs from the GOES response.

A more direct comparison with GOES can be made using the Solar Assembly for X-rays (SAX) aboard the MErcury Surface, Space ENvironment, GEochemistry, and Ranging (MESSENGER) satellite. SAX, like GOES XRS, also measures soft X-ray flux directly, albeit over a different disc-integrated view and a slightly different energy range. Spectral fitting to the SAX data can be used to derive the flare temperature and emission measure, which can in turn be used to calculate theoretical GOES fluxes for the flare. Neither of these relatively simple methods for deriving a GOES proxy from the MESSENGER data have been examined, so we do so here.

While the primary goal of this study is to determine a relationship between SAX and XRS, a potential side-benefit is the discovery of occulted flare candidates, where SAX sees the whole flare but an imaging-capable instrument observes purely coronal emission.
Occulted flares can be valuable diagnostic sources for faint coronal emission, especially where X-ray imaging spectrometers such as Reuven Ramaty High Energy Solar Spectroscopic Imager (RHESSI; \citet{linReuvenRamatyHighEnergy2002}) are concerned. Due to RHESSI's limited dynamic range, coronal emission can only be viewed unobstructed when the X-ray bright chromospheric footpoints are hidden behind the solar limb. There are several well-studied examples of occulted events, which often take the form of arcades with looptop and over-the-looptop sources (e.g. \citet{masudaLooptopHardXray1994},  \citet{kruckerHardXrayEmission2008}).
In cases of extreme occultation, so-called double coronal sources can even be observed well into the low-density regime high above the flare loop arcade (\citet{chenDoubleCoronalXray2017}, \citet{lastufkaMultiwavelengthStereoscopicObservation2019}).

Using RHESSI, \citet{kruckerHardXrayEmission2008} identified a list of 55 occulted flares during the maximum of solar cycle 23. Recently, \citet{effenbergerHardXRayEmission2017} expanded the list to include solar cycle 24. A list of 398 occulted flare candidates observed by RHESSI has been compiled by
 M. Oka\footnote{Full list available at: \url{http://www.ssl.berkeley.edu/\~moka/rhessi/flares\_occulted.html}}.
  In events like these, GOES does not observe the full extent of the SXR emission.
 Thus, the MESSENGER data provides two possible improvements in the realm of occulted flare studies. Firstly, relating SAX data directly to GOES class can constrain the estimate of total energy for the 600+ events observed by MESSENGER.
Secondly, the MESSENGER data can confirm the presence of a significant soft X-ray source for those occulted flare candidates with favorable geometry. 

In this work, we first empirically derived the relationship between MESSENGER data and GOES flux (section \ref{sec:direct}. We used this relation to determine the proxy for determining the GOES class of a solar flare, based on MESSENGER data. In section \ref{sec:calc}, we checked how the GOES-equivalent flux derived from SAX spectra compared to what XRS actually measured, which revealed intrinsic differences between the two soft X-ray instruments. We then compared both the MESSENGER flux to the STEREO flux, where available, and the derived MESSENGER proxy with the STEREO/GOES relation obtained by \citet{nittaSoftXrayFluxes2013}, in section \ref{sec:euv}. Finally, in section \ref{sec:flares}, we used RHESSI imaging to identify MESSENGER events that would have been highly occulted to an Earth-orbiting instrument.

\section{Soft X-ray Flux Relationship}\label{sec:flux}

The SAX was part of the XRS (X-ray spectrometer) aboard the Mercury-orbiting MESSENGER satellite. It measured full-disk integrated soft X-rays between 1.5 and 8.5 keV with a resolution of 598 eV at 5.9 keV \citep{boyntonMarsOdysseyGammaRay2004}. The majority of the data was collected over 5 minute intervals, but the instrument was capable of triggering a high-cadence mode of 40 s or 20 s during the most intense flares. 
\citet{dennisSOLARFLAREELEMENT2015} filtered the data for events where a two-thermal fit to the lightcurves resulted in 
temperatures between 0.5 and 2.5 keV, 
emission measures between 0.01 and 2.5 \rm{x} $10^{49}$ cm$^{-3}$ and a reduced chi-squared between 0.5 and 1.8.\footnote{The original paper also stated a selection criterion of more than 5000 counts per second in the 6.3--7 keV bin, but this was never applied in the actual routine used for analysis (private communication, \citet{dennisClarificationMESSENGERFlare2018})}
This resulted in a list of 657 flares that occurred between May 28, 2007 and  August 19, 2013. Due to a period of low solar activity, there were no flares recorded by SAX between January 2008 and February 2010. 

GOES satellites 14 and 15 were operational during the MESSENGER mission, with 14 acting as the on-orbit spare to back up 15. Spatially integrated X-ray flux was measured every three seconds, and data was almost always available. The GOES wavelength ranges correspond to energies of 3--25 keV and 1.5--12.4 keV for the short and long channels respectively.

Other instruments available at the time which are relevant to this study were the Solar Dynamics Observatory's Atmospheric Imager Assembly (AIA), operational since April 2010, which provided additional imaging for all but the 50 earliest events on the MESSENGER list. Both STEREO Ahead and Behind data were available from May 2007. RHESSI, operational since 2002, was a source of hard X-ray imaging spectroscopy as its operational cycle allowed.

\begin{figure}
\includegraphics[width=.5\textwidth]{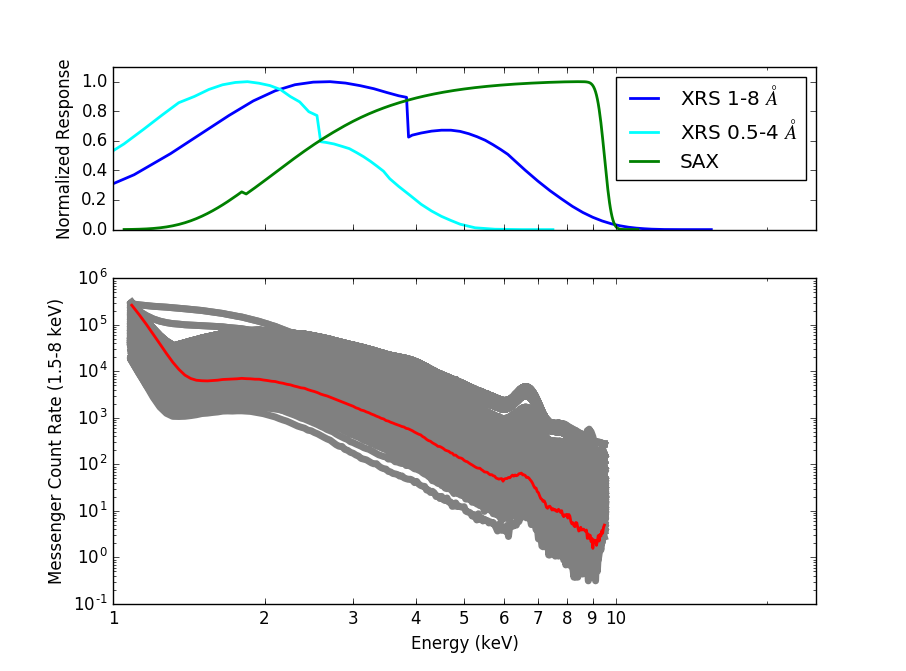}
\caption{The soft X-ray instrument normalized response functions (above) and the MESSENGER data set of $\sim$650 flares (below), represented by the spectra at the flare peak. The flare of 1 June, 2007, which is shown in additional detail in Figure 1 of \citet{dennisSOLARFLAREELEMENT2015}, is highlighted in red. The range from 2.5--7.5 keV is used in Section \ref{sec:calc} to fit the MESSENGER spectra.}
\label{fig:all_MESSENGER}
\end{figure}

\subsection{Direct comparison of MESSENGER and GOES measurements}\label{sec:direct}

By far the simplest method of comparison between the MESSENGER SAX and GOES XRS datasets is in their native units. SAX data, after being corrected for the difference in light travel time, is presented in terms of binned counts over a certain time interval. GOES XRS data is first converted to units of flux ($\rm{Wm^{-2}}$). To further simplify matters, no pre-flare background subtraction was performed during this analysis, for either instrument. 

We identified the flare peak for each MESSENGER event by looking for the count maximum recorded in the upper energy channels, $3.2$--$5.3$ and $5.3$--$8.5$ keV, in the twenty minutes following the recorded onset time given by \citet{dennisSOLARFLAREELEMENT2015}. 
From this peak time, we extracted two different count values, each chosen to correspond best to either the GOES short or GOES long energy range. In other words, the sum of all counts occurring in the energy range from 3--10 keV (SAX itself has a nominal energy range of ~1--10 keV) was used to compare to GOES 0.5--4 \AA{} measurements, while the sum of SAX counts from 1.5--10 keV was used to compare to GOES 1--8 \AA{} data. 
These total count maxima were then divided by the duration of the time bin in which they were recorded -- normally 300 s.
The range of peak counts per second for the MESSENGER flare list had a minimum of 10$^2$ and a maximum of 2 \rm{x} 10$^4$ in both ranges. 
We chose the units of counts per second in the corresponding energy ranges as the most elementary means of comparison. Both the OSPEX GUI and the paper of \citet{dennisSOLARFLAREELEMENT2015} use counts $\rm{s}^{-1} \rm{keV}^{-1} \rm{cm}^{-2}$, although the units of the binary files where the data is stored is simply counts in a given time and energy bin. 

These count maxima were compared to the GOES flux in the relevant channel, averaged over the MESSENGER peak time bin. For this reason, the GOES flux values used here are not the same ones (the peak value of the event lightcurve) that are used to determine GOES class.

\begin{figure}
\includegraphics[width=.5\textwidth]{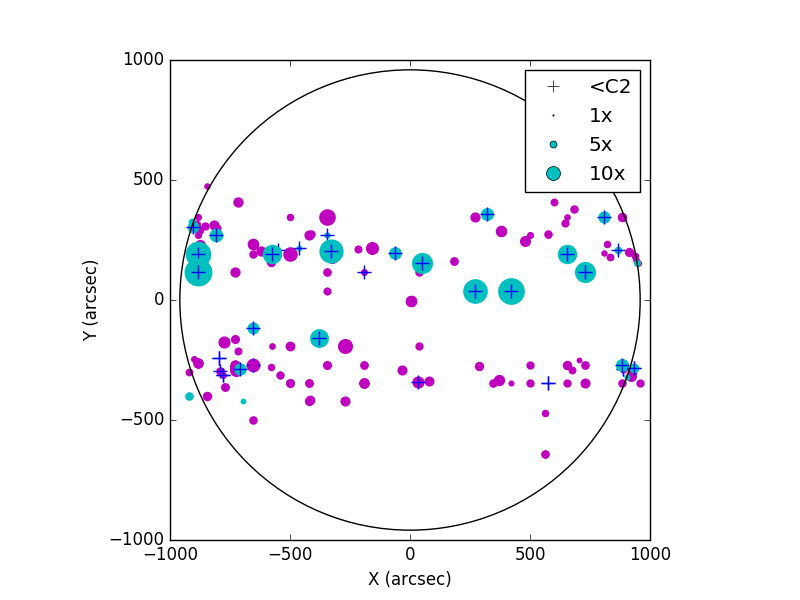}
\caption{Locations of flares as given by HEK/AIA. Flares are color-coded according to the ratio of the theoretical GOES flux calculated from the MESSENGER temperature and emission measure to the actual GOES flux (see Figure \ref{fig:calc} -- MESSENGER/GOES $<$ 1 is cyan and $>$1 is magenta) and the size of the symbol associated with them is directly proportional to the value of the ratio. Events marked with a plus sign indicate those where the measured GOES flux is less than class C2, corresponding to the red triangles in Figure \ref{fig:badflares}.} 
\label{fig:aia_loc}
\end{figure}

To determine if an event in the MESSENGER flare list was potentially observable by GOES, we made use of the Heliophysics Event Knowledgebase (HEK; \citet{hurlburtHeliophysicsEventKnowledgebase2012}).
First, we checked for GOES or AIA events which occurred within five minutes of the MESSENGER peak time interval and were classified as flares. If such an event was present, the coordinates of that flare were used to determine if the region in question was visible or not to MESSENGER, given its position in Mercury orbit. Figure \ref{fig:aia_loc} shows the AIA-recorded locations for all 299 jointly visible events. 
The locations of flare occurrence are evenly divided between the two hemispheres and follows the typical distribution of magnetic flux the Sun. 

\begin{figure}
\includegraphics[width=.5\textwidth]{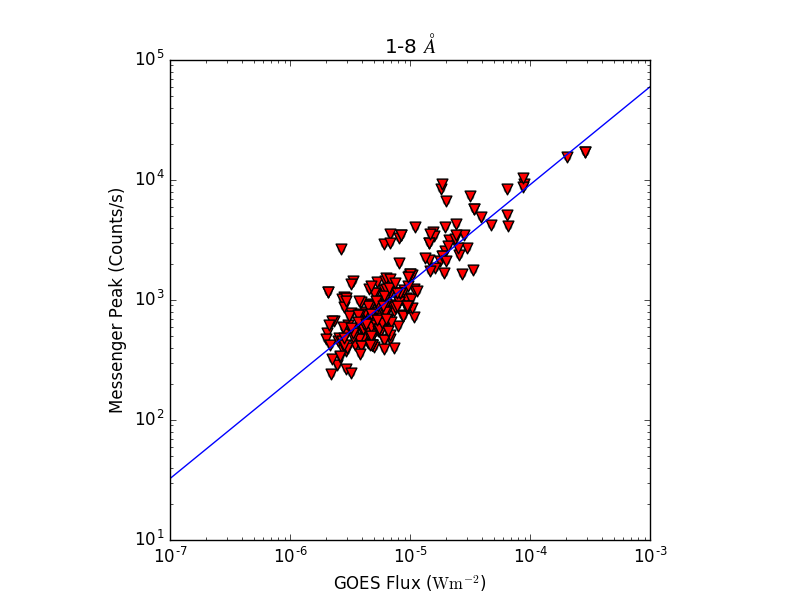}
\includegraphics[width=.5\textwidth]{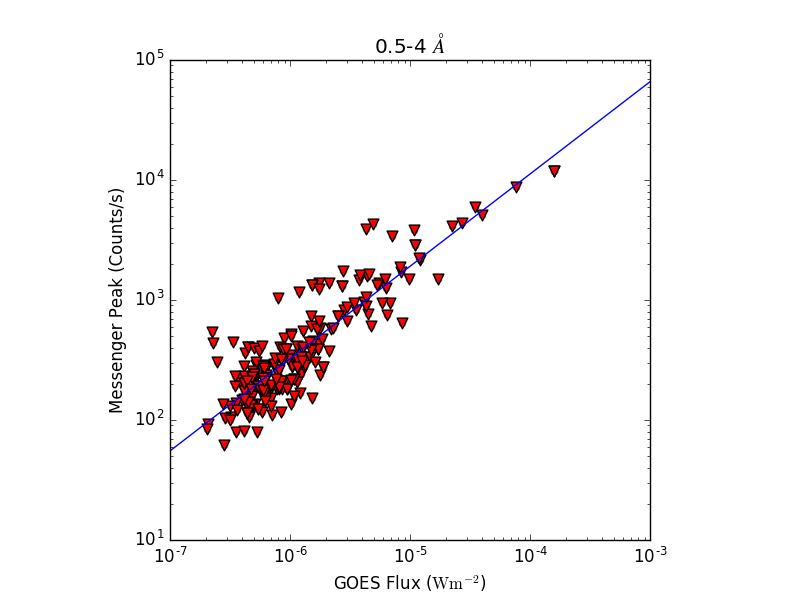}
\caption{Relationship of MESSENGER peak flux to GOES peak flux. Events marked in red are the flares visible to both satellites. The blue shows the line of best fit above a lower threshold (C2 for GOES long, B2 for GOES short).}
\label{fig:direct}
\vspace*{\floatsep}
\includegraphics[width=.5\textwidth]{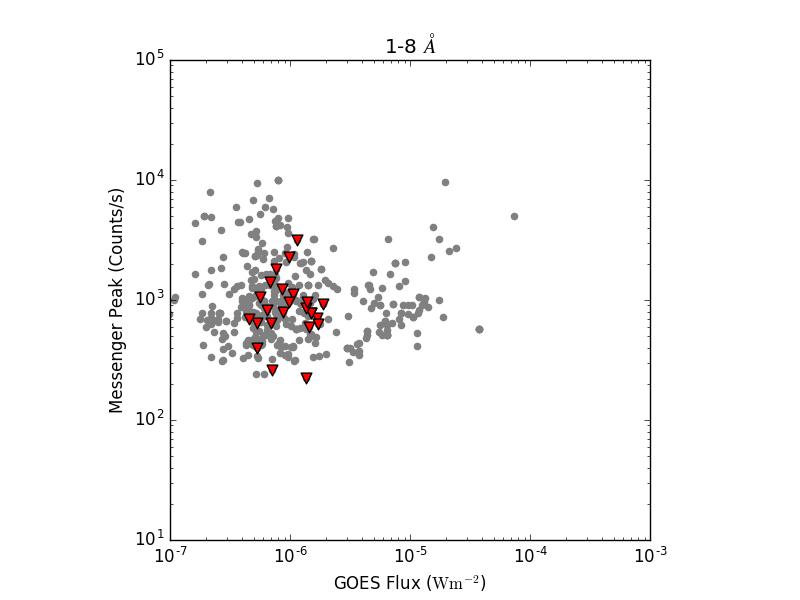}
\caption{Events which were excluded from the samples above in Figure \ref{fig:direct}. Although jointly visible to MESSENGER and GOES, events marked in red were deemed unsuitable for other reasons, described in the text.}
\label{fig:badflares}
\end{figure}

Figure \ref{fig:direct} shows the comparison between the MESSENGER peak counts per second and the average GOES flux of these flares, for both MESSENGER count summation ranges and both GOES channels.
The correlation between MESSENGER peak counts and GOES flux is strong (R = 0.75) above a certain lower GOES class, with the threshold extending lower for the short wavelength channel. 
For the 0.5--4 \AA{} channel, we chose a lower boundary of 2\rm{x}10$^{-7} \rm{W m}^{-2}$, or B2, while an order of magnitude higher was required for the 1--8 \AA{} channel (C2).
The resulting lines of best fit are shown in blue in Figure \ref{fig:direct}.

 Letting C represent the peak counts per second measured by MESSENGER, either in the low- or high-energy ranges previously described, and F the GOES flux (W m$^{-2}$) in the specified channel, the fits are given by:

\begin{equation}\label{eq:Mproxy} 
F_{GOES(long)}  = 1.40 \pm0.19 \rm{x}10^{-9} * C_{MESSENGER(low)}^{1.22\pm.04}
\end{equation}

and

\begin{equation}
F_{GOES(short)} = 4.91 \pm0.17 \rm{x} 10^{-10}* C_{MESSENGER(high)}^{1.31\pm.03}
\end{equation}

From here onward, we will refer to GOES fluxes calculated using Equation \ref{eq:Mproxy} as the MESSENGER proxy values.

The data display clear power-law relationships with low error on the exponents. 
Notably, the exponents are both greater than one. This aligns with findings in \citet{warmuthConstraintsEnergyRelease2016}, which along with past studies demonstrates how higher temperature flares emit more flux in the GOES wavelengths than lower temperature flares do. 
The data in Figure \ref{fig:direct} lie, on average, within 50\% (GOES long) and 80\% (GOES short) of the value determined by the fits. This is comparable to what was found in the Nitta and Chertok studies.
 
For completeness, Figure \ref{fig:badflares} shows flares that were excluded from the fitting range for one of two reasons: either the event was not deemed jointly visible by both instruments (grey dots) or it was likely misidentified by the aforementioned selection method (red triangles). There were twenty-six flares that fell under the latter situation. These all were below measured GOES class C2 and most had a comparatively much higher number of MESSENGER counts than was expected, given the fit to the rest of the data shown in the two accompanying plots. The positions of these flares are marked with additional plus signs on Figure \ref{fig:aia_loc}. These events were at first considered candidates for occulted flares; however, a close examination of the outliers revealed that their high SAX-to-GOES ratios were due to exceptional situations rather than MESSENGER observing the event on-disk while GOES sees purely coronal emission.
Seven of these events occurred between 28 -- 30 October 2011, a very active period with a high ($\sim$10$^-6$ W m$^{-2}$, or C-class) GOES background. 
Three events (SOL2011-10-18T23:45, SOL2012-06-29T01:21 and SOL2011-06-30T12:49) showed similar rises in the GOES and MESSENGER lightcurves, however with the MESSENGER count rates reaching a maximum in the five-minute time bin before the GOES rise had truly begun. Because the time interval used to measure average GOES flux was based on the time of the MESSENGER peak, the result was a higher number of counts relative to flux than would be obtained if using the five-minute period surrounding the true local maximum in both curves. Even though some of these outliers were located near the solar limb, none of them represented a situation where GOES measured only coronal emission from an occulted flare. In searching for occulted events from this flare list, therefore, we must consider all the flares, not just those with high MESSENGER counts to GOES flux ratios.

\subsection{GOES flux derived from isothermal fits to MESSENGER spectra}\label{sec:calc}

Given the temperature and emission measure of a plasma, one can calculate the expected GOES flux \citep{whiteUpdatedExpressionsDetermining2005}. By deriving temperature and emission measure from the MESSENGER spectra, we are able to compare expected GOES flux to the GOES flux that was actually measured. 

The MESSENGER spectra can be well-fit (reduced chi-squared of $\sim$0.5) at higher energies ($>$2.5 -- 7.5 keV)\footnote{The 7.5 keV cutoff was necessary to avoid contamination of the spectra at higher energies which occurs when charged flare particles at these energies interact in the detector.}, allowing us to derive temperature and emission measures for each flare. We used the Object Spectral Executive (OSPEX) package\footnote{Documentation at: \url{https://hesperia.gsfc.nasa.gov/ssw/packages/spex/doc/ospex_explanation.htm}} to do this for all MESSENGER events, assuming a single-temperature plasma model and leaving the fraction of the Chianti solar elemental abundances as a free parameter in the fit\footnote{\citet{dennisSOLARFLAREELEMENT2015} used a second fit to better accommodate the low-energy spectrum, and also varied the instrumental response to better fit the Fe line complex.}.
Finally, with the \textit{goes\_fluxes.pro} routine, we calculated the predicted GOES flux in each channel. 

\begin{figure}
\includegraphics[width=.5\textwidth]{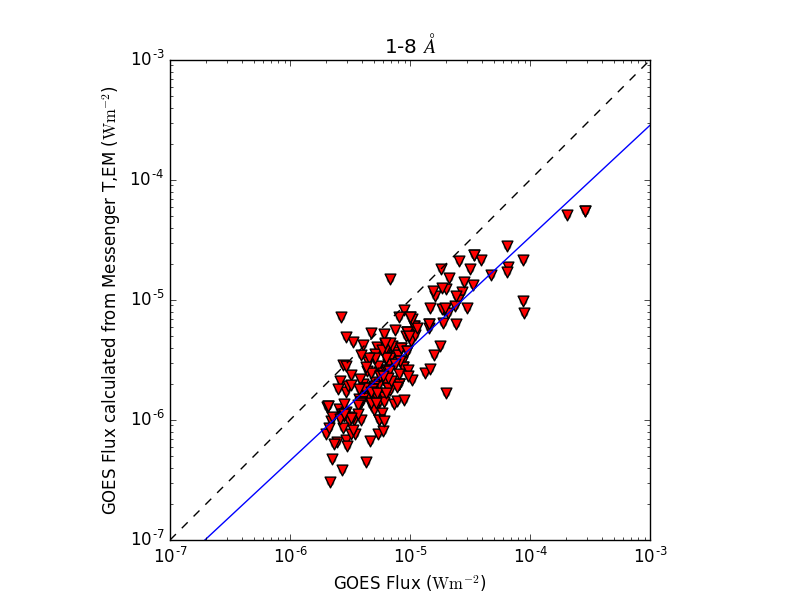}
\includegraphics[width=.5\textwidth]{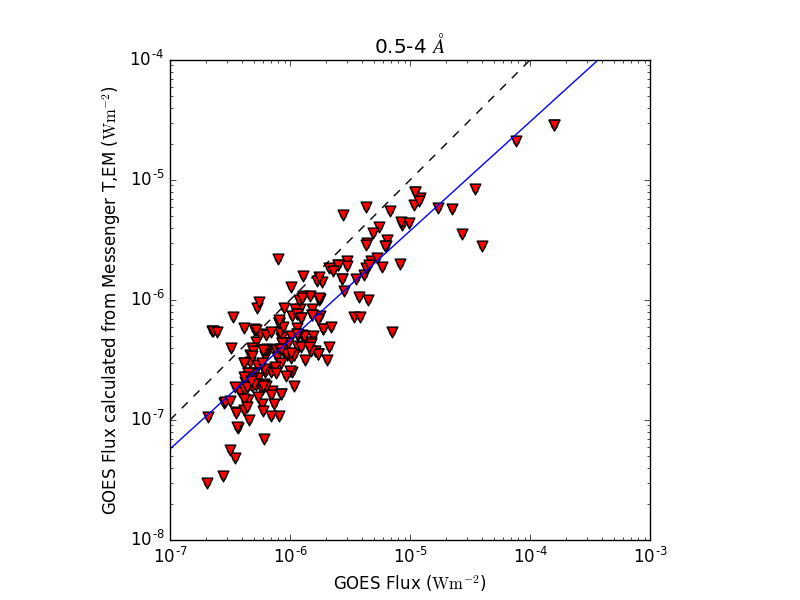}
\caption{Relationship of GOES-equivalent flux derived from the MESSENGER spectra to the measured GOES flux, for all jointly observed flares above a certain flux threshold (C2 for GOES long, B2 for GOES short). The blue shows the line of best fit, while the dashed black line is the 1:1 ratio.}
\label{fig:calc}
\end{figure}

Figure \ref{fig:calc} shows that there is a strong correlation between the calculated and measured GOES fluxes. While the slope of the best fit to the data, especially in the long wavelength channel, which has more overlap with the MESSENGER energy range, is close to one, the fit nevertheless results in underestimation of the GOES-equivalent flux by about a factor of two. 

\begin{figure}
\includegraphics[width=.5\textwidth]{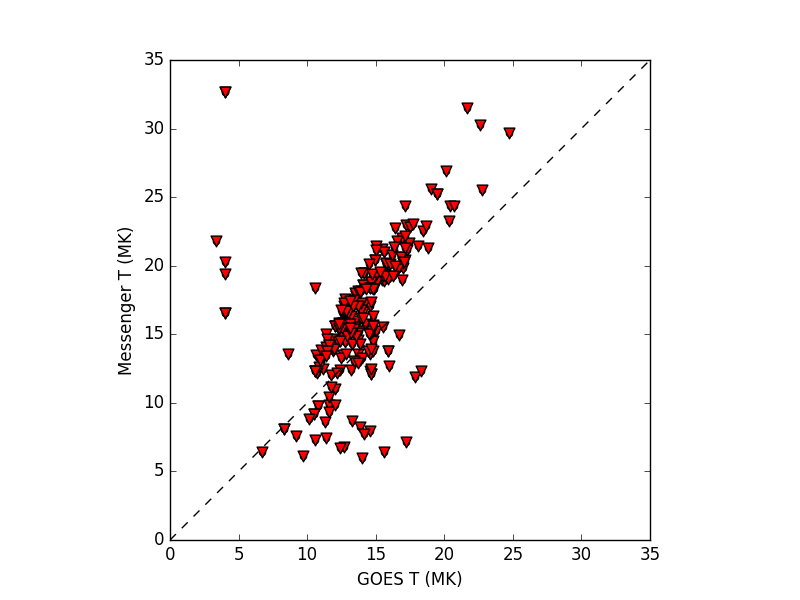}
\includegraphics[width=.5\textwidth]{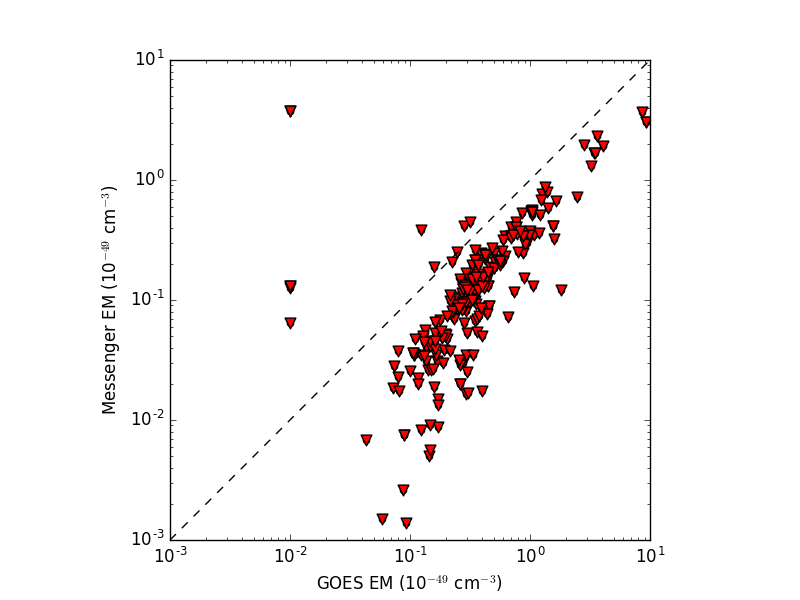}
\caption{Temperatures and emission measures calculated for each instrument. MESSENGER values were obtained via OSPEX fit, while GOES quantities were calculated using the SSWIDL routine \textit{goes\_teem.pro}, based on empirical data.}
\label{fig:tem}
\end{figure}

Using the same lower limits on the GOES flux as in Section \ref{sec:direct} and excluding the two highest points, which significantly change the result of the fit, we use $F_{MESSENGER}$ to represent the calculated SXR flux and obtain the following relationships:

\begin{equation}
F_{GOES(long)} = 6.36\pm0.23 * F_{MESSENGER(long)}^{1.07\pm .05} 
\end{equation}

and

\begin{equation}\label{eq:Mproxy2}
F_{GOES(short)} = 18.86\pm0.27 * F_{MESSENGER(short)}^{1.15\pm .04} 
\end{equation}

The reasons for the disagreement between SAX and XRS are a the result of the complexity of joint observations. Firstly, consider the instrument response functions (top plot of Figure \ref{fig:all_MESSENGER}), which show the energy ranges where the instruments are most sensitive, at different energies. 
In addition to the instrumental response, one must consider the dominant emission mechanism at these wavelengths. Thermal bremsstrahlung spectra strongly depend on plasma temperature, and the two instruments have different temperature biases, with MESSENGER being sensitive to  hotter plasma than GOES.

If plasma filling flare loops was at a single uniform temperature, this disparity in sampled energy range wouldn't matter. However, flares are not isothermal, and low temperature plasma, emitting low-energy soft X-rays, contributes the most to a multithermal plasma's emission measure. Therefore, if all the flares observed contain a low-temperature component contributing significantly to the emission measure, a fit to the high-energy MESSENGER spectrum will miss this contribution. That is indeed what we see in Figure \ref{fig:tem}; MESSENGER underestimates flare emission measures by a similar factor to which the GOES-equivalent flux is underestimated.

\section{Comparison with EUV}\label{sec:euv}

Estimates of the magnitude of flares hidden from the earth-orbiting GOES satellites have been done using STEREO in the EUV, by \citet{nittaSoftXrayFluxes2013} and \citet{chertokSimpleWayEstimate2015}. The Nitta method is more versatile as it does not rely on the presence of a saturation streak, so we chose to use this method for comparison. 

Nitta's empirical formula, derived from a sample of over 400 $>$C3 flares observed by GOES and STEREO, is:

\begin{equation}\label{eq:nitta}
F_{GOES (long)}= 1.39 \rm{x} 10^{-11}\it{F_{EUVI(195)}} 
\end{equation}

$F_{EUVI(195)}$ is determined by subtracting a full-disk, calibrated pre-flare 195 \AA{} image from a likewise-prepared image at the flare peak. Units of these images are not in DNs$^{-1}$ (data numbers per second) as originally stated in their work. 
The correct units are those output by \emph{secchi\_prep}, which in its default mode is photons/s. In the 195 \AA{} channel, 1 DN/s = 0.862 photons/s; therefore, a small correction factor can recover the relation in DN/s (private communication, \citet{nittaClarificationUnitSecchi2018, liuClarificationUnitSecchi2018}). 

Their derived empirical fit was best for flares with a flux over $4$\rm{x}$10^{6}$ 
photons/s, with an estimated spread of 0.5--1.5$F_{EUVI(195)}$ and 0.3-3 $F_{EUVI(195)}$ for lower fluxes. This flux, according to their derived fit, corresponds to a GOES class of M5.5. Because the MESSENGER data set did not contain many flares of M-class or larger, we do not expect to find a 
strong relationship between the GOES X-ray and the STEREO EUV fluxes for our set of mainly medium sized flares, which are also expected to be much cooler than M-class or larger flares (see for example \citet{warmuthConstraintsEnergyRelease2016}'s Figure 3). 

Simply subtracting the full disk images in the way specified by this method has some drawbacks. Contamination from CCD snow produced by CMEs or particle hits can artificially increase the calculated difference, resulting in an overestimate of the GOES class. 
As suggested by Liu \citep{liuEUVIFluxCorrection2018}, who calculated this proxy for work in \citet{effenbergerHardXRayEmission2017}, the best practice would be to select a field of view including the flare itself but small relative to the solar disk to minimize the impacts of contamination. 

Nevertheless, we used the original \citet{nittaSoftXrayFluxes2013} method to calculate the EUV fluxes and resulting GOES estimate for all 118 flares determined to be jointly visible by MESSENGER, GOES, and one of the STEREO spacecraft. Joint visibility was calculated using flare coordinates from the STEREO EUVI event database for events not already registered with a location in the HEK database.  

\begin{figure}
\includegraphics[width=.5\textwidth]{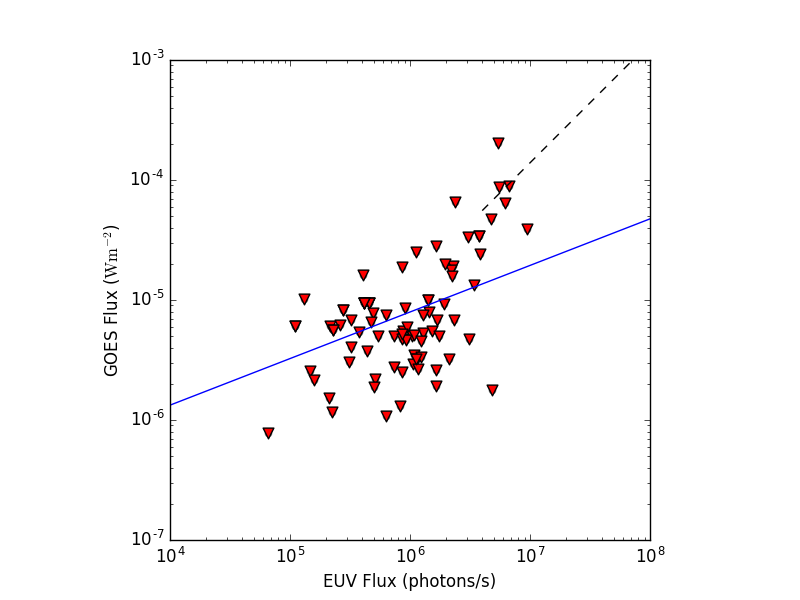} 
\includegraphics[width=.5\textwidth]{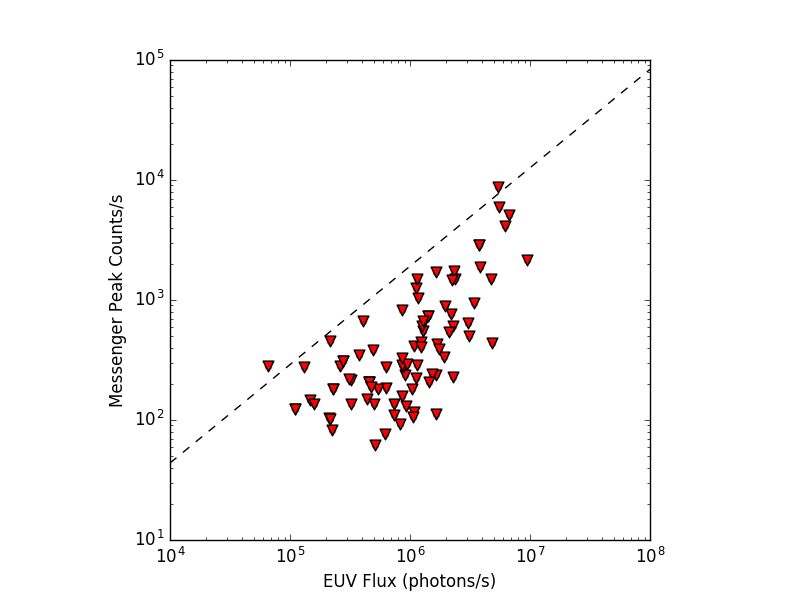} 
\includegraphics[width=.5\textwidth]{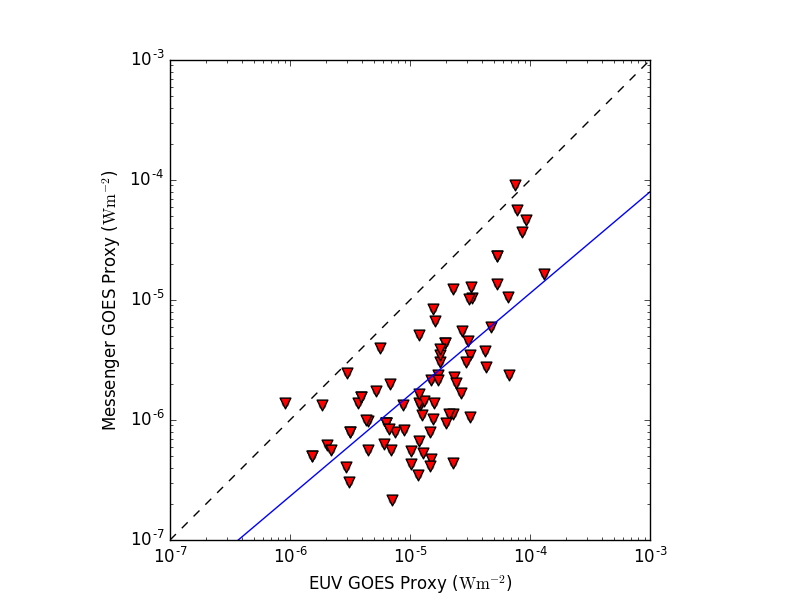} 
\caption{Comparison of SXR fluxes with EUV flux. Lines of best fit are shown in blue. In the top plot, the dashed line indicates the fit found in \citet{nittaSoftXrayFluxes2013} and begins at the value above which the fit is best (GOES M5.5 or 4x10$^6$ photons/s in EUV 195 \AA). In the middle plot, it illustrates the derived MESSENGER proxy (Equation \ref{eq:Mproxy}), after conversion via the EUV-GOES proxy (Equation \ref{eq:nitta}). In the bottom plot, the dashed line indicates the 1:1 ratio.}
\label{fig:NFall}
\end{figure}

We used an image from one hour before the MESSENGER peak time as the pre-flare image. The results are shown in Figure \ref{fig:NFall}. The top panel shows a direct comparison with the GOES 1--8 \AA{} flux measured for each event, while the middle shows the comparison to the MESSENGER peak counts. Finally, the last panel shows a comparison between the empirically derived proxies for both MESSENGER (equation \ref{eq:Mproxy}) and the EUV (equation \ref{eq:nitta}).

Eighty-six of these events had a GOES long-wavelength flux above class C3, the aforementioned lower limit of events included in the Nitta study. 
Comparing measured GOES flux to the calculated EUV proxy, the data show a good positive correlation (R=0.55) that is close to the strength of the correlation for the entire population of flares studied by Nitta (R=0.67). 
A correlation between MESSENGER peak counts and the EUV flux difference is also evident (R=0.67). 
The correlation itself is barely improved when separating events by temperature, although the standard error of the linear regression decreases by half when considering only flares with temperatures above 25 MK. However, the already small sample size decreases accordingly. 

Comparing the two proxies allows us to see the areas in which one method might lead to a very different GOES flux estimate than the other. The bottom panel of Figure \ref{fig:NFall} clearly shows that the EUV proxy gives values up to one GOES class greater than those of the SAX proxy. Although there are only 7 flares greater than M5 --- the value above which the Nitta proxy works best --- in this sample, they do show a different trend than the rest of the data. A fit through these points still shows SAX underestimating the GOES flux relative to the EUV, but by a factor of 3 instead of almost 10. 

The disparity in the two proxies is well illustrated by looking at the three  major behind-the-limb (as seen from GOES) events listed by \citep{nittaSoftXrayFluxes2013} that were also observed by MESSENGER. These flares, on 31 August and 1 September 2010 and 23 October 2011, are given GOES classifications of X1--X2 by Nitta. Chertok's method of measuring the saturation streak of the EUV peak image gave similar or more conservative estimates for each of these events, while the MESSENGER counts led to estimates up to an order of magnitude lower than those of the Nitta method (see Table \ref{tbl:ratios}). For each of these events, the time bin containing the MESSENGER peak counts was within about a minute of the STEREO peak image. It is unclear why there is still such a large discrepancy between the two proxies, even though both STEREO and MESSENGER have close relationships with GOES flux for events of this magnitude. 

\begin{deluxetable*}{cccccc}
\tabletypesize{\scriptsize}
\tablecaption{Comparison between GOES proxies for large flares (W m$^{-2}$)\label{tbl:ratios}}
\tablewidth{0pt}
\tablehead{
\colhead{Date} & \colhead{Time} &\colhead{Location (STEREO)}&\colhead{Nitta} & \colhead{Chertov} & \colhead{MESSENGER}}
\startdata
2010-08-31&20:55:53&(A) S22 W146&1.7*10$^{-4}$&1.42*10$^{-4}$&2.30*10$^{-5}$\\ 
2010-09-01&21:50:53&(A) S22W162&1.1*10$^{-4}$&1.14*10$^{-4}$&3.84*10$^{-5}$\\ 
2011-10-23&23:15:44&(A) N19W151&1.1*10$^{-4}$&7.77*10$^{-5}$&2.07*10$^{-5}$ 
\enddata

\end{deluxetable*}

\section{Occulted flares observed by MESSENGER and RHESSI}\label{sec:flares}

The MESSENGER dataset provides an opportunity to identify occulted 
flares that have not been previously noticed. Of particular interest are events that were at least partially visible to RHESSI, which would provide valuable X-ray imaging spectroscopy of coronal sources. The RHESSI flare detection algorithm requires a source to appear in each image reconstructed by the three coarsest subcollimators before it is flagged as a flare. It is possible that faint coronal sources associated with a highly occulted flare might escape the notice of the RHESSI flare identification routines. The HEK flare detection algorithm is likewise based on images, namely heavily binned AIA data 
that is searched for areas of positive derivative and negative derivative around a local maximum \citep{martensComputerVisionSolar2012}. 

Either way, faint, extended X-ray sources above the limb, or their EUV counterparts, might not produce enough emission at the required spatial scales to trigger a flare alert. 
Therefore, an excess of soft X-ray signal from one viewing angle with respect to the Earth's could indicate the presence of a flare that is only partially visible to Earth. We examined every event in the MESSENGER dataset with a GOES flux less than 2 \rm{x} $10^{-6}$ (class C2), as well as every event whose HEK catalog position was close to the solar limb, as an occulted flare candidate.

To identify true coronal sources, we ultimately needed RHESSI imaging observations. We checked the selected flare time intervals for RHESSI observations, which resulted in the elimination of 49 events where data was unavailable. For flares with quick-look images available, we excluded events that were on the disk. The remaining time intervals were used to generate full-disk RHESSI back-projection images using subcollimators 8 and 9 in the 6--25 keV range, during the 116 seconds surrounding the event peak. The flare location was taken to be at the coordinates of the brightest point in the image. 
For events with high counting statistics, finer resolution RHESSI imaging, done using CLEAN, was used to derive a more precise source centroid position. 
Examining these images finally allowed us to select flares that were jointly observed, with $>10^{4}$ counts in the 12--25 keV range and beyond-the-limb emission visible by RHESSI. 

The final list of 20 events is given in Table \ref{tbl:flares}. Of these, only four were noted in previous works. The STEREO-given flare location is listed if available in the EUVI database, otherwise the most likely location of the flaring active region was determined for STEREO images. Images from two events, observed by MESSENGER during very active periods on the Sun, did not show a noticeably flaring region, so no location information is given for them. GOES class is listed from the HEK catalog where available; otherwise, the values from this study -- the average GOES flux during the MESSENGER peak time bin -- are used. These values are noticeably smaller than catalog values, because they are averaged over a five-minute period. The MESSENGER proxy values from Equation \ref{eq:Mproxy}, are given for comparison, as well as an estimate of the occultation angle, using the AIA limb for calculation.

\begin{deluxetable*}{lllcccc}
\tabletypesize{\scriptsize}
\tablecaption{Occulted flares observed by RHESSI and MESSENGER.}
\tablewidth{0pt}
\tablehead{
\colhead{MESSENGER} & \colhead{Location} & \colhead{Location} &\colhead{RHESSI}& \colhead{GOES} &\colhead{MESSENGER}&\colhead{Occultation} \\
\colhead{Peak Date} & \colhead{RHESSI} & \colhead{STEREO}&\colhead{Max Energy}&\colhead{Class}&\colhead{Proxy}&\colhead{angle}\\
\colhead{} & \colhead{x,y arcsec} & \colhead{x,y arcsec}&\colhead{(keV)}&\colhead{(catalog)}&\colhead{(GOES class)}&\colhead{(deg)}}
\startdata
2007-05-30 13:46:33&	[-944,-128]	& B, [-79, -3, 0.984] & 20 & C2.2 & M4.3& \\
2010-11-03 12:11:43$^{1}$&	[-912,-336]	& B, [-225,-391] & 50 & C4.9 & M2.7&40\\
2011-09-12 05:51:45&	[944,240]	& A, [-206, 294] & 20 & C2.3 & C7.9&3\\ 
2011-09-18 23:54:43&	[864,464]	& A & 10 & C1.0* & M5.8&\\ 
2011-10-22 04:33:32&	[-944,96]	& A, [901,273] & 15 & C1.4 & C9.2&71\\ 
2011-12-16 00:32:42&	[928,-320]	& B, [94,-200] & 10 & B3.6* & C5.0&40\\ 
2012-01-15 05:14:58&	[-928,288]	& A, [423,422] & 15 & B8.7* & M1.0&77 \\ 
2012-03-24 08:41:09&	[-896,-416]	& B, [542,-410] & 10 & C7.2 & M5.6&78\\ 
2012-05-05 05:10:29&	[-928,192]	&A, [862,-226] & 20 & B7.7* & M1.2&68\\ 
2012-06-25 18:15:09&	[-944,-256]& B, [192,-128] & 15 & B5.4 & C7.2&71\\ 
2012-07-19 05:10:01$^{2,4}$& [960,-240]& A, [-479,-276] & 100 & M7.8  & M6.0&8\\
2012-09-12 11:10:07&	[976,-256]&	B, [-617,-305] & 10 & B9.1 & X1.1&29\\ 
2012-09-12 22:50:03&	[990,-230]	& B, [-542,-301] & 10 & B7.1 & M8.9&21\\ 
2012-09-17 06:17:45&	[-976,224]	& B, [191,-131] & 15 & B5.6 & C3.9&54\\ 
2012-10-17 07:54:16$^{3}$&	[-960,112]	& B, [807,-371] & 30 & C7.4 & M1.5&84\\ 
2012-10-20 18:07:40$^{3,4}$&	[-960,-240]	& B, [450,-212] & 40 & M8.8 & X1.6&73\\ 
2012-10-21 05:28:05&	[-944,-256]	& B, [532,-226] & 20 & C5.6 & C6.7&76\\ 
2013-01-28 20:12:15&	[-944,128]	&A, [934,146] & 10 & B3.1* & M1.0&48\\ 
2013-04-18 18:07:16&	[976,192]	&A, [-770, 111] & 20 & C6.5 & C5.9&7\\ 
2013-04-27 10:49:49&   [964,205]  &A & 20 & B9.1* & C3.9
\enddata
\label{tbl:flares}
\tablenotetext{1}{Studied by \citet{glesenerObservationHeatingFlareaccelerated2013}}
\tablenotetext{2}{Listed by \citet{kruckerParticleDensitiesAcceleration2014,sunDifferentialEmissionMeasure2014,wuMicrowaveImagingHot2016}}
\tablenotetext{3}{Listed by \citet{effenbergerHardXRayEmission2017}}
\tablenotetext{4}{Studied by \citet{kruckerCoSpatialWhiteLight2015}}
\tablenotetext{d}{RHESSI sees decay phase only}
\tablenotetext{*}{Not from catalog - from GOES lightcurve}
\end{deluxetable*}

\section{Summary} 

GOES class has long been the accepted metric of flare classification. As such, a proxy is useful in order to fit events which occur out of sight of GOES into the typical schema used to compare flares. 
In this work we showed how the peak counts of the MESSENGER Solar Assembly for X-rays (SAX) can be used to determine the equivalent GOES class of a flare using a simple relation. For flares above class C2, we recovered an empirical power-law relationship, although with an uncertainty which on average creates a scatter of a factor of two. A similar relationship with the short-wavelength GOES channel can be determined down to class B2. 
Therefore, SAX data can be used to determine equivalent GOES class for a large number of moderate- to large-sized flares that occur on the far side of the Sun as observed by GOES.

It is tempting to expect very similar results from SAX and XRS, since they are both soft X-ray instruments. A comparison of the expected GOES flux, based on fits to the MESSENGER spectra, to what XRS actually measured, warns against such simplifying assumptions about both the instruments and about the thermal nature of flares. The data clearly showed that MESSENGER, sensitive to the higher energy range, predicted lower fluxes in the GOES wavelength channels than were actually measured. This would not happen if the X-ray emitting plasma was isothermal. 

The multithermal nature of flares can be easy to overlook, especially when a one- or two-temperature function is often sufficient to obtain a good fit to an X-ray spectrum.
DEM analysis using observations from many wavelengths is a good illustration of the many temperature components that constitute thermal emission from flare plasma \citep[e.g.][]{ishikawaDetectionNanoflareheatedPlasma2017,wrightMicroflareHeatingSolar2017}. 
Although it is most commonly undertaken with EUV observations, carefully cross-calibrated hard and soft X-ray data can additionally be used to find out what temperature plasma dominates emission at certain wavelengths \citep[e.g.][]{mctiernanMultiInstrumentEVERHESSISolar2015}. 
Even disregarding the possibility that low (5--6 keV) energy emissions might in fact be due to non-thermal processes, these studies are very informative. 
A good example with purely thermal emission is \citet{ishikawaHotPlasmaQuiescent2019}, who visualize contributions of different temperature components to active region emission in Figure 6 of their work. Lower energy emission is due to lower temperature plasma; for example, the DEM of RHESSI between 3.5 and 4.5 keV shows that almost half of the emission is due to plasma with temperatures below 5 MK. Compare this with a higher energy, 6.5--7.5 keV (where SAX is most sensitive), where this low-temperature plasma does not contribute at all to the DEM, and most of the emission is in fact due to plasma with temperatures 10 MK and above. With this picture in mind, it becomes clear that SAX and GOES observe different temperature components of flares.

The multithermal nature of solar flare plasmas also affects proxies derived from narrowband EUV filter observations. Although its response function is double peaked at $\sim$ 2 and $\sim$20 MK, emission at 195 \AA{} is still dominated by cooler $\approx$1 MK plasma, except during large solar flares.
A well-known proxy derived from EUV image data was found by \citet{nittaSoftXrayFluxes2013}, using the difference in the peak- and pre-flare 195 \AA{} images. 
Comparing the results of the two proxies, using SAX data as derived in this work and using STEREO images as per \cite{nittaSoftXrayFluxes2013}, we found that the MESSENGER proxy gave GOES class estimates up to an order of magnitude lower than the EUV-derived values. 
It is important to keep in mind, however, that the population of flares examined here is likely much cooler on average than the ones used to derive the EUV proxy. 
Given the large scatter of the data in Figure \ref{fig:NFall}, especially for smaller flares, we recommend use of the SAX proxy derived in section \ref{sec:direct} when SAX data is available. For flares the high-temperature component dominates emission at 195 \AA, the EUV proxy would give results within the same range of accuracy (factor of 2).

With the examples of these two proxies to draw upon, it is already clear that empirically deriving a GOES proxy is an exercise fraught with subtle complications. 
Even the Miniature X-ray Solar Spectrometer (MinXSS, \cite{mooreInstrumentsCapabilitiesMiniature2018}), covering a very similar energy range as the GOES long channel, demonstrated that the XRS-MinXSS relationship between disk- and band-integrated flux was non-linear \citep{woodsNewSolarIrradiance2017}. Yet proxies for the GOES X-ray flux will likely be sought after for years to come, especially with the increasing number of non-Earth orbiting instruments. 
Missions to the inner heliosphere will rely on proxies for determining GOES classes. Although SAX is no longer operational, the recently launched Mercury mission BepiColombo carries the Solar Intensity X-ray and particle Spectrometer (SIXS) \citep{huovelinSolarIntensityXray2010}, with a slightly broader energy range but similar scientific purpose to SAX.
Observing at energies up to 150 keV, the Spectrometer/Telescope for Imaging X-rays (STIX, \citet{kruckerSpectrometerTelescopeImaging2020}, aboard Solar Orbiter \citep{mullerSolarOrbiter2013}, will experience similar difficulties to RHESSI; that is, a large bias towards high-temperature plasma (see \citet{warmuthConstraintsEnergyRelease2016} Figure 3). 
This study clearly demonstrates the importance of considering biases that may arise when comparing instruments that potentially observe different temperature components of the same flare.

 \section*{Acknowledgements}\label{sec:ack}

The work was supported by Swiss National Science Foundation (200021-163377). The authors would like to thank Brian Dennis, whose comments improved the paper.

\bibliographystyle{plainnat}
\bibliography{Zotero.bib}

\begin{thebibliography}{37}
\providecommand{\natexlab}[1]{#1}
\providecommand{\url}[1]{\texttt{#1}}
\expandafter\ifx\csname urlstyle\endcsname\relax
  \providecommand{\doi}[1]{doi: #1}\else
  \providecommand{\doi}{doi: \begingroup \urlstyle{rm}\Url}\fi

\bibitem[Baker(1970)]{bakerFlareClassificationBased1970}
Donald~M. Baker.
\newblock Flare classification based upon {{X}}-ray intensity.
\newblock \emph{The American Institute of Aeronautics and Astronautics (AIAA)},
  1970.

\bibitem[Boynton et~al.(2004)Boynton, Feldman, Mitrofanov, Evans, Reedy,
  Squyres, Starr, Trombka, D'uston, Arnold, Englert, Metzger, W{\"a}nke,
  Br{\"u}ckner, Drake, Shinohara, Fellows, Hamara, Harshman, Kerry, Turner,
  Ward, Barthe, Fuller, Storms, Thornton, Longmire, Litvak, and
  Ton'chev]{boyntonMarsOdysseyGammaRay2004}
W.~V. Boynton, W.~C. Feldman, I.~G. Mitrofanov, L.~G. Evans, R.~C. Reedy, S.~W.
  Squyres, R.~Starr, J.~I. Trombka, C.~D'uston, J.~R. Arnold, P.~A.~J. Englert,
  A.~E. Metzger, H.~W{\"a}nke, J.~Br{\"u}ckner, D.~M. Drake, C.~Shinohara,
  C.~Fellows, D.~K. Hamara, K.~Harshman, K.~Kerry, C.~Turner, M.~Ward,
  H.~Barthe, K.~R. Fuller, S.~A. Storms, G.~W. Thornton, J.~L. Longmire, M.~L.
  Litvak, and A.~K. Ton'chev.
\newblock The {{Mars Odyssey Gamma}}-{{Ray Spectrometer Instrument Suite}}.
\newblock In Christopher~T. Russell, editor, \emph{2001 {{Mars Odyssey}}},
  pages 37--83. {Springer Netherlands}, {Dordrecht}, 2004.
\newblock ISBN 978-0-306-48600-5.

\bibitem[Chen et~al.(2017)Chen, Wu, Liu, Schwartz, Zhao, Wang, and
  Du]{chenDoubleCoronalXray2017}
Yao Chen, Zhao Wu, Wei Liu, Richard~A. Schwartz, Di~Zhao, Bing Wang, and Guohui
  Du.
\newblock Double {{Coronal X}}-ray and {{Microwave Sources Associated With A
  Magnetic Breakout Solar Eruption}}.
\newblock \emph{The Astrophysical Journal}, 843\penalty0 (1):\penalty0 8, June
  2017.
\newblock ISSN 1538-4357.
\newblock \doi{10.3847/1538-4357/aa7462}.

\bibitem[Chertok et~al.(2015)Chertok, Belov, and
  Grechnev]{chertokSimpleWayEstimate2015}
Ilya Chertok, A.V. Belov, and V.V. Grechnev.
\newblock A simple way to estimate the soft x-ray class of far-side solar
  flares observed with {{STEREO EUVI}}.
\newblock \emph{Solar Physics}, May 2015.

\bibitem[Dennis(21.08.18)]{dennisClarificationMESSENGERFlare2018}
Brian~R. Dennis.
\newblock Clarification of {{Messenger}} flare list selection criteria,
  21.08.18.

\bibitem[Dennis et~al.(2015)Dennis, Phillips, Schwartz, Tolbert, Starr, and
  Nittler]{dennisSOLARFLAREELEMENT2015}
Brian~R. Dennis, Kenneth J.~H. Phillips, Richard~A. Schwartz, Anne~K. Tolbert,
  Richard~D. Starr, and Larry~R. Nittler.
\newblock Solar {{Flare Element Abundances}} from the {{Solar Assembly}} for
  {{X}}-{{Rays}} ({{SAX}}) on {{MESSENGER}}.
\newblock \emph{The Astrophysical Journal}, 803:\penalty0 67, April 2015.
\newblock ISSN 0004-637X.
\newblock \doi{10.1088/0004-637X/803/2/67}.

\bibitem[Donnelly(1979)]{donnellySolarActivityReports1979}
Richard~Frank Donnelly.
\newblock \emph{Solar Activity Reports}.
\newblock {Department of Commerce, National Oceanic and Atmospheric
  Administration, Environmental Research Laboratories}, 1979.

\bibitem[Effenberger et~al.(2017)Effenberger, da~Costa, Oka, {Saint-Hilaire},
  Liu, {Vah\'e Petrosian}, Glesener, and
  Krucker]{effenbergerHardXRayEmission2017}
Frederic Effenberger, Fatima~Rubio da~Costa, Mitsuo Oka, Pascal
  {Saint-Hilaire}, Wei Liu, {Vah\'e Petrosian}, Lindsay Glesener, and S{\"a}m
  Krucker.
\newblock Hard {{X}}-{{Ray Emission}} from {{Partially Occulted Solar Flares}}:
  {{RHESSI Observations}} in {{Two Solar Cycles}}.
\newblock \emph{The Astrophysical Journal}, 835\penalty0 (2):\penalty0 124,
  2017.
\newblock ISSN 0004-637X.
\newblock \doi{10.3847/1538-4357/835/2/124}.

\bibitem[Glesener et~al.(2013)Glesener, Krucker, Bain, and
  Lin]{glesenerObservationHeatingFlareaccelerated2013}
Lindsay Glesener, S{\"a}m Krucker, Hazel~M. Bain, and Robert~P. Lin.
\newblock Observation of {{Heating}} by {{Flare}}-accelerated {{Electrons}} in
  a {{Solar Coronal Mass Ejection}}.
\newblock \emph{The Astrophysical Journal Letters}, 779\penalty0 (2):\penalty0
  L29, 2013.
\newblock ISSN 2041-8205.
\newblock \doi{10.1088/2041-8205/779/2/L29}.

\bibitem[Hannah et~al.(2011)Hannah, Hudson, Battaglia, Christe, Ka{\v
  s}parov{\'a}, Krucker, Kundu, and
  Veronig]{hannahMicroflaresStatisticsXray2011}
I.~G. Hannah, H.~S. Hudson, M.~Battaglia, S.~Christe, J.~Ka{\v s}parov{\'a},
  S.~Krucker, M.~R. Kundu, and A.~Veronig.
\newblock Microflares and the {{Statistics}} of {{X}}-ray {{Flares}}.
\newblock \emph{Space Science Reviews}, 159:\penalty0 263--300, September 2011.
\newblock ISSN 0038-6308.
\newblock \doi{10.1007/s11214-010-9705-4}.

\bibitem[Howard et~al.(2008)Howard, Moses, Vourlidas, Newmark, Socker,
  Plunkett, Korendyke, Cook, Hurley, Davila, Thompson, St~Cyr, Mentzell,
  Mehalick, Lemen, Wuelser, Duncan, Tarbell, Wolfson, Moore, Harrison, Waltham,
  Lang, Davis, Eyles, {Mapson-Menard}, Simnett, Halain, Defise, Mazy, Rochus,
  Mercier, Ravet, Delmotte, Auchere, Delaboudiniere, Bothmer, Deutsch, Wang,
  Rich, Cooper, Stephens, Maahs, Baugh, McMullin, and
  Carter]{howardSunEarthConnection2008}
R.~A. Howard, J.~D. Moses, A.~Vourlidas, J.~S. Newmark, D.~G. Socker, S.~P.
  Plunkett, C.~M. Korendyke, J.~W. Cook, A.~Hurley, J.~M. Davila, W.~T.
  Thompson, O.~C. St~Cyr, E.~Mentzell, K.~Mehalick, J.~R. Lemen, J.~P. Wuelser,
  D.~W. Duncan, T.~D. Tarbell, C.~J. Wolfson, A.~Moore, R.~A. Harrison, N.~R.
  Waltham, J.~Lang, C.~J. Davis, C.~J. Eyles, H.~{Mapson-Menard}, G.~M.
  Simnett, J.~P. Halain, J.~M. Defise, E.~Mazy, P.~Rochus, R.~Mercier, M.~F.
  Ravet, F.~Delmotte, F.~Auchere, J.~P. Delaboudiniere, V.~Bothmer, W.~Deutsch,
  D.~Wang, N.~Rich, S.~Cooper, V.~Stephens, G.~Maahs, R.~Baugh, D.~McMullin,
  and T.~Carter.
\newblock Sun {{Earth Connection Coronal}} and {{Heliospheric Investigation}}
  ({{SECCHI}}).
\newblock \emph{Space Science Reviews}, 136:\penalty0 67--115, April 2008.
\newblock ISSN 0038-6308.
\newblock \doi{10.1007/s11214-008-9341-4}.

\bibitem[Huovelin et~al.(2010)Huovelin, Vainio, Andersson, Valtonen, Alha,
  M{\"a}lkki, Grande, Fraser, Kato, Koskinen, Muinonen, N{\"a}r{\"a}nen,
  Schmidt, Syrj{\"a}suo, Anttila, Vihavainen, Kiuru, Roos, Peltonen, Lehti,
  Talvioja, Portin, and Prydderch]{huovelinSolarIntensityXray2010}
J.~Huovelin, R.~Vainio, H.~Andersson, E.~Valtonen, L.~Alha, A.~M{\"a}lkki,
  M.~Grande, G.~W. Fraser, M.~Kato, H.~Koskinen, K.~Muinonen,
  J.~N{\"a}r{\"a}nen, W.~Schmidt, M.~Syrj{\"a}suo, M.~Anttila, T.~Vihavainen,
  E.~Kiuru, M.~Roos, J.~Peltonen, J.~Lehti, M.~Talvioja, P.~Portin, and
  M.~Prydderch.
\newblock Solar {{Intensity X}}-ray and particle {{Spectrometer}} ({{SIXS}}).
\newblock \emph{Planetary and Space Science}, 58\penalty0 (1):\penalty0
  96--107, January 2010.
\newblock ISSN 0032-0633.
\newblock \doi{10.1016/j.pss.2008.11.007}.

\bibitem[Hurlburt et~al.(2012)Hurlburt, Cheung, Schrijver, Chang, Freeland,
  Green, Heck, Jaffey, Kobashi, Schiff, Serafin, Seguin, Slater, Somani, and
  Timmons]{hurlburtHeliophysicsEventKnowledgebase2012}
N.~Hurlburt, M.~Cheung, C.~Schrijver, L.~Chang, S.~Freeland, S.~Green, C.~Heck,
  A.~Jaffey, A.~Kobashi, D.~Schiff, J.~Serafin, R.~Seguin, G.~Slater,
  A.~Somani, and R.~Timmons.
\newblock Heliophysics {{Event Knowledgebase}} for~the~{{Solar Dynamics
  Observatory}} ({{SDO}}) and {{Beyond}}.
\newblock \emph{Solar Physics}, 275\penalty0 (1):\penalty0 67--78, January
  2012.
\newblock ISSN 1573-093X.
\newblock \doi{10.1007/s11207-010-9624-2}.

\bibitem[Ishikawa and Krucker(2019)]{ishikawaHotPlasmaQuiescent2019}
Shin-nosuke Ishikawa and S{\"a}m Krucker.
\newblock Hot {{Plasma}} in a {{Quiescent Solar Active Region}} as {{Measured}}
  by {{RHESSI}}, {{XRT}}, and {{AIA}}.
\newblock \emph{The Astrophysical Journal}, 876\penalty0 (2):\penalty0 111, May
  2019.
\newblock ISSN 0004-637X.
\newblock \doi{10.3847/1538-4357/ab13a1}.

\bibitem[Ishikawa et~al.(2017)Ishikawa, Glesener, Krucker, Christe,
  {Buitrago-Casas}, Narukage, and
  Vievering]{ishikawaDetectionNanoflareheatedPlasma2017}
Shin-nosuke Ishikawa, Lindsay Glesener, S{\"a}m Krucker, Steven Christe,
  Juan~Camilo {Buitrago-Casas}, Noriyuki Narukage, and Juliana Vievering.
\newblock Detection of nanoflare-heated plasma in the solar corona by the
  {{FOXSI}}-2 sounding rocket.
\newblock \emph{Nature Astronomy}, 1:\penalty0 771--774, October 2017.
\newblock ISSN 2397-3366.
\newblock \doi{10.1038/s41550-017-0269-z}.

\bibitem[Krucker et~al.(2008)Krucker, Battaglia, Cargill, Fletcher, Hudson,
  MacKinnon, Masuda, Sui, Tomczak, Veronig, Vlahos, and
  White]{kruckerHardXrayEmission2008}
S.~Krucker, M.~Battaglia, P.~J. Cargill, L.~Fletcher, H.~S. Hudson, A.~L.
  MacKinnon, S.~Masuda, L.~Sui, M.~Tomczak, A.~L. Veronig, L.~Vlahos, and S.~M.
  White.
\newblock Hard {{X}}-ray emission from the solar corona.
\newblock \emph{The Astronomy and Astrophysics Review}, 16\penalty0
  (3):\penalty0 155--208, December 2008.
\newblock ISSN 1432-0754.
\newblock \doi{10.1007/s00159-008-0014-9}.

\bibitem[Krucker et~al.(2020)Krucker, Hurford, Grimm, K{\"o}gl,
  Gr{\"o}belbauer, Etesi, Casadei, Csillaghy, Benz, Arnold, Molendini,
  Orleanski, Schori, Xiao, Kuhar, Kobler, Iseli, Dreier, Wiehl, and
  Kleint]{kruckerSpectrometerTelescopeImaging2020}
S.~Krucker, G.~J. Hurford, O.~Grimm, S.~K{\"o}gl, H.-P. Gr{\"o}belbauer,
  L.~Etesi, D.~Casadei, A.~Csillaghy, A.~O. Benz, N.~G. Arnold, F.~Molendini,
  P.~Orleanski, D.~Schori, H.~Xiao, M.~Kuhar, S.~Kobler, L.~Iseli, M.~Dreier,
  H.~J. Wiehl, and L.~Kleint.
\newblock The {{Spectrometer}}/{{Telescope}} for {{Imaging X}}-rays ({{STIX}}).
\newblock \emph{Astronomy \& Astrophysics}, January 2020.
\newblock ISSN 0004-6361, 1432-0746.
\newblock \doi{10.1051/0004-6361/201937362}.

\bibitem[Krucker and Battaglia(2014)]{kruckerParticleDensitiesAcceleration2014}
S{\"a}m Krucker and Marina Battaglia.
\newblock Particle {{Densities}} within the {{Acceleration Region}} of a
  {{Solar Flare}}.
\newblock \emph{The Astrophysical Journal}, 780\penalty0 (1):\penalty0 107,
  2014.
\newblock ISSN 0004-637X.
\newblock \doi{10.1088/0004-637X/780/1/107}.

\bibitem[Krucker et~al.(2015)Krucker, Hilaire, Hudson, Haberreiter, Oliveros,
  Fivian, Hurford, Kleint, Battaglia, Kuhar, and
  Arnold]{kruckerCoSpatialWhiteLight2015}
S{\"a}m Krucker, Pascal~Saint Hilaire, Hugh~S. Hudson, Margit Haberreiter, Juan
  Carlos~Martinez Oliveros, Martin~D. Fivian, Gordon Hurford, Lucia Kleint,
  Marina Battaglia, Matej Kuhar, and Nicolas~G. Arnold.
\newblock Co-{{Spatial White Light}} and {{Hard X}}-{{Ray Flare Footpoints Seen
  Above}} the {{Solar Limb}}.
\newblock \emph{The Astrophysical Journal}, 802\penalty0 (1):\penalty0 19,
  2015.
\newblock ISSN 0004-637X.
\newblock \doi{10.1088/0004-637X/802/1/19}.

\bibitem[Lastufka et~al.(2019)Lastufka, Krucker, Zimovets, Nizamov, White,
  Masuda, Golovin, Litvak, Mitrofanov, and
  Sanin]{lastufkaMultiwavelengthStereoscopicObservation2019}
Erica Lastufka, S{\"a}m Krucker, Ivan Zimovets, Bulat Nizamov, Stephen White,
  Satoshi Masuda, Dmitriy Golovin, Maxim Litvak, Igor Mitrofanov, and Anton
  Sanin.
\newblock Multiwavelength {{Stereoscopic Observation}} of the 2013 {{May}} 1
  {{Solar Flare}} and {{CME}}.
\newblock \emph{The Astrophysical Journal}, 886\penalty0 (1):\penalty0 9,
  November 2019.
\newblock ISSN 0004-637X.
\newblock \doi{10.3847/1538-4357/ab4a0a}.

\bibitem[Lin et~al.(2002)Lin, Dennis, Hurford, Smith, Zehnder, Harvey, Curtis,
  Pankow, Turin, Bester, Csillaghy, Lewis, Madden, {van Beek}, Appleby,
  Raudorf, McTiernan, Ramaty, Schmahl, Schwartz, Krucker, Abiad, Quinn, Berg,
  Hashii, Sterling, Jackson, Pratt, Campbell, Malone, Landis,
  {Barrington-Leigh}, {Slassi-Sennou}, Cork, Clark, Amato, Orwig, Boyle, Banks,
  Shirey, Tolbert, Zarro, Snow, Thomsen, Henneck, McHedlishvili, Ming, Fivian,
  Jordan, Wanner, Crubb, Preble, Matranga, Benz, Hudson, Canfield, Holman,
  Crannell, Kosugi, Emslie, Vilmer, Brown, {Johns-Krull}, Aschwanden, Metcalf,
  and Conway]{linReuvenRamatyHighEnergy2002}
R.~P. Lin, B.~R. Dennis, G.~J. Hurford, D.~M. Smith, A.~Zehnder, P.~R. Harvey,
  D.~W. Curtis, D.~Pankow, P.~Turin, M.~Bester, A.~Csillaghy, M.~Lewis,
  N.~Madden, H.~F. {van Beek}, M.~Appleby, T.~Raudorf, J.~McTiernan, R.~Ramaty,
  E.~Schmahl, R.~Schwartz, S.~Krucker, R.~Abiad, T.~Quinn, P.~Berg, M.~Hashii,
  R.~Sterling, R.~Jackson, R.~Pratt, R.~D. Campbell, D.~Malone, D.~Landis,
  C.~P. {Barrington-Leigh}, S.~{Slassi-Sennou}, C.~Cork, D.~Clark, D.~Amato,
  L.~Orwig, R.~Boyle, I.~S. Banks, K.~Shirey, A.~K. Tolbert, D.~Zarro, F.~Snow,
  K.~Thomsen, R.~Henneck, A.~McHedlishvili, P.~Ming, M.~Fivian, John Jordan,
  Richard Wanner, Jerry Crubb, J.~Preble, M.~Matranga, A.~Benz, H.~Hudson,
  R.~C. Canfield, G.~D. Holman, C.~Crannell, T.~Kosugi, A.~G. Emslie,
  N.~Vilmer, J.~C. Brown, C.~{Johns-Krull}, M.~Aschwanden, T.~Metcalf, and
  A.~Conway.
\newblock The {{Reuven Ramaty High}}-{{Energy Solar Spectroscopic Imager}}
  ({{RHESSI}}).
\newblock \emph{Solar Physics}, 210:\penalty0 3--32, November 2002.
\newblock ISSN 0038-0938.
\newblock \doi{10.1023/A:1022428818870}.

\bibitem[Liu(03.08.18)]{liuEUVIFluxCorrection2018}
Wei Liu.
\newblock {{EUVI}} flux correction for behind-the-limb flares, 03.08.18.

\bibitem[Liu(14.08.18)]{liuClarificationUnitSecchi2018}
Wei Liu.
\newblock Clarification of the unit of secchi\_prep output, 14.08.18.

\bibitem[Martens et~al.(2012)Martens, Attrill, Davey, Engell, Farid, Grigis,
  Kasper, Korreck, Saar, Savcheva, Su, Testa, {Wills-Davey}, Bernasconi,
  Raouafi, Delouille, Hochedez, Cirtain, DeForest, Angryk, De~Moortel,
  Wiegelmann, Georgoulis, McAteer, and Timmons]{martensComputerVisionSolar2012}
P.~C.~H. Martens, G.~D.~R. Attrill, A.~R. Davey, A.~Engell, S.~Farid, P.~C.
  Grigis, J.~Kasper, K.~Korreck, S.~H. Saar, A.~Savcheva, Y.~Su, P.~Testa,
  M.~{Wills-Davey}, P.~N. Bernasconi, N.-E. Raouafi, V.~A. Delouille, J.~F.
  Hochedez, J.~W. Cirtain, C.~E. DeForest, R.~A. Angryk, I.~De~Moortel,
  T.~Wiegelmann, M.~K. Georgoulis, R.~T.~J. McAteer, and R.~P. Timmons.
\newblock Computer {{Vision}} for the {{Solar Dynamics Observatory}} ({{SDO}}).
\newblock \emph{Solar Physics}, 275\penalty0 (1):\penalty0 79--113, January
  2012.
\newblock ISSN 1573-093X.
\newblock \doi{10.1007/s11207-010-9697-y}.

\bibitem[Masuda et~al.(1994)Masuda, Kosugi, Hara, Tsuneta, and
  Ogawara]{masudaLooptopHardXray1994}
S.~Masuda, T.~Kosugi, H.~Hara, S.~Tsuneta, and Y.~Ogawara.
\newblock A loop-top hard {{X}}-ray source in a compact solar flare as evidence
  for magnetic reconnection.
\newblock \emph{Nature}, 371\penalty0 (6497):\penalty0 495--497, October 1994.
\newblock ISSN 1476-4687.
\newblock \doi{10.1038/371495a0}.

\bibitem[McTiernan et~al.(2015)McTiernan, Caspi, and
  Warren]{mctiernanMultiInstrumentEVERHESSISolar2015}
James~M. McTiernan, Amir Caspi, and Harry Warren.
\newblock The {{Multi}}-{{Instrument}} ({{EVE}}-{{RHESSI}}) {{DEM}} for {{Solar
  Flares}}, and {{Implications}} for {{Residual Non}}-{{Thermal Soft X}}-{{Ray
  Emission}}.
\newblock \emph{l}, 1:\penalty0 302.10, April 2015.

\bibitem[Moore et~al.(2018)Moore, Caspi, Woods, Chamberlin, Dennis, Jones,
  Mason, Schwartz, and Tolbert]{mooreInstrumentsCapabilitiesMiniature2018}
Christopher~S. Moore, Amir Caspi, Thomas~N. Woods, Phillip~C. Chamberlin,
  Brian~R. Dennis, Andrew~R. Jones, James~P. Mason, Richard~A. Schwartz, and
  Anne~K. Tolbert.
\newblock The {{Instruments}} and {{Capabilities}} of the {{Miniature X}}-{{Ray
  Solar Spectrometer}} ({{MinXSS}}) {{CubeSats}}.
\newblock \emph{Solar Physics}, 293\penalty0 (2):\penalty0 21, January 2018.
\newblock ISSN 1573-093X.
\newblock \doi{10.1007/s11207-018-1243-3}.

\bibitem[M{\"u}ller et~al.(2013)M{\"u}ller, Marsden, Cyr, Gilbert, and
  Team]{mullerSolarOrbiter2013}
D.~M{\"u}ller, R.~G. Marsden, O.~C.~St Cyr, H.~R. Gilbert, and The
  Solar~Orbiter Team.
\newblock Solar {{Orbiter}}.
\newblock \emph{Solar Physics}, 285\penalty0 (1-2):\penalty0 25--70, July 2013.
\newblock ISSN 0038-0938, 1573-093X.
\newblock \doi{10.1007/s11207-012-0085-7}.

\bibitem[Nitta et~al.(2013)Nitta, Aschwanden, Boerner, Freeland, Lemen, and
  Wuelser]{nittaSoftXrayFluxes2013}
N.~V. Nitta, M.~J. Aschwanden, P.~F. Boerner, S.~L. Freeland, J.~R. Lemen, and
  J.-P. Wuelser.
\newblock Soft {{X}}-ray {{Fluxes}} of {{Major Flares Far Behind}} the {{Limb}}
  as {{Estimated Using STEREO EUV Images}}.
\newblock \emph{Solar Physics}, 288\penalty0 (1):\penalty0 241--254, November
  2013.
\newblock ISSN 0038-0938, 1573-093X.
\newblock \doi{10.1007/s11207-013-0307-7}.

\bibitem[Nitta(14.08.18)]{nittaClarificationUnitSecchi2018}
Nariaki Nitta.
\newblock Clarification of the unit of secchi\_prep output, 14.08.18.

\bibitem[Ryan et~al.(2012)Ryan, Milligan, Gallagher, Dennis, Tolbert, Schwartz,
  and Young]{ryanThermalPropertiesSolar2012}
Daniel~F. Ryan, Ryan~O. Milligan, Peter~T. Gallagher, Brian~R. Dennis, A.~Kim
  Tolbert, Richard~A. Schwartz, and C.~Alex Young.
\newblock The {{Thermal Properties}} of {{Solar Flares Over Three Solar Cycles
  Using GOES X}}-ray {{Observations}}.
\newblock \emph{The Astrophysical Journal Supplement Series}, 202\penalty0
  (2):\penalty0 11, October 2012.
\newblock ISSN 0067-0049, 1538-4365.
\newblock \doi{10.1088/0067-0049/202/2/11}.

\bibitem[Sun et~al.(2014)Sun, Cheng, and
  Ding]{sunDifferentialEmissionMeasure2014}
J.~Q. Sun, X.~Cheng, and M.~D. Ding.
\newblock Differential emission measure {{Analysis}} of a limb solar flare on
  2012 {{July}} 19.
\newblock \emph{The Astrophysical Journal}, 786\penalty0 (1):\penalty0 73,
  April 2014.
\newblock ISSN 0004-637X, 1538-4357.
\newblock \doi{10.1088/0004-637X/786/1/73}.

\bibitem[Warmuth and Mann(2016)]{warmuthConstraintsEnergyRelease2016}
A.~Warmuth and G.~Mann.
\newblock Constraints on energy release in solar flares from {{RHESSI}} and
  {{GOES X}}-ray observations - {{I}}. {{Physical}} parameters and scalings.
\newblock \emph{Astronomy \& Astrophysics}, 588:\penalty0 A115, April 2016.
\newblock ISSN 0004-6361, 1432-0746.
\newblock \doi{10.1051/0004-6361/201527474}.

\bibitem[White et~al.(2005)White, Thomas, and
  Schwartz]{whiteUpdatedExpressionsDetermining2005}
Stephen~M. White, Roger~J. Thomas, and Richard~A. Schwartz.
\newblock Updated {{Expressions}} for {{Determining Temperatures}} and
  {{Emission Measures}} from {{Goes Soft X}}-{{Ray Measurements}}.
\newblock \emph{Solar Physics}, 227:\penalty0 231--248, April 2005.
\newblock \doi{10.1007/s11207-005-2445-z}.

\bibitem[Woods et~al.(2017)Woods, Caspi, Chamberlin, Jones, Kohnert, Mason,
  Moore, Palo, Rouleau, Solomon, Machol, and
  Viereck]{woodsNewSolarIrradiance2017}
Thomas~N. Woods, Amir Caspi, Phillip~C. Chamberlin, Andrew Jones, Richard
  Kohnert, James~Paul Mason, Christopher~S. Moore, Scott Palo, Colden Rouleau,
  Stanley~C. Solomon, Janet Machol, and Rodney Viereck.
\newblock New {{Solar Irradiance Measurements}} from the {{Miniature X}}-{{Ray
  Solar Spectrometer CubeSat}}.
\newblock \emph{ApJ}, 835\penalty0 (2):\penalty0 122, February 2017.
\newblock ISSN 0004-637X.
\newblock \doi{10.3847/1538-4357/835/2/122}.

\bibitem[Wright et~al.(2017)Wright, Hannah, Grefenstette, Glesener, Krucker,
  Hudson, Smith, Marsh, White, and Kuhar]{wrightMicroflareHeatingSolar2017}
Paul~J. Wright, Iain~G. Hannah, Brian~W. Grefenstette, Lindsay Glesener,
  S{\"a}m Krucker, Hugh~S. Hudson, David~M. Smith, Andrew~J. Marsh, Stephen~M.
  White, and Matej Kuhar.
\newblock Microflare {{Heating}} of a {{Solar Active Region Observed}} with
  {{NuSTAR}}, {{Hinode}}/{{XRT}}, and {{SDO}}/{{AIA}}.
\newblock \emph{The Astrophysical Journal}, 844:\penalty0 132, August 2017.
\newblock ISSN 0004-637X.
\newblock \doi{10.3847/1538-4357/aa7a59}.

\bibitem[Wu et~al.(2016)Wu, Chen, Huang, Nakajima, Song, Melnikov, Liu, Li,
  Chandrashekhar, and Jiao]{wuMicrowaveImagingHot2016}
Z.~Wu, Y.~Chen, G.~Huang, H.~Nakajima, H.~Song, V.~Melnikov, W.~Liu, G.~Li,
  K.~Chandrashekhar, and F.~Jiao.
\newblock Microwave imaging of a hot flux rope structure during the
  pre-impulsive stage of an eruptive {{M7}}.7 solar flare.
\newblock \emph{The Astrophysical Journal}, 820\penalty0 (2):\penalty0 L29,
  March 2016.
\newblock ISSN 2041-8213.
\newblock \doi{10.3847/2041-8205/820/2/L29}.

\end{thebibliography}

\end{document}